\documentclass[lettersize,journal]{IEEEtran}
\usepackage{amsmath,amssymb,amsfonts}
\usepackage{algorithmic}
\usepackage{algorithm}
\usepackage{array}
\usepackage{textcomp}
\usepackage{stfloats}
\usepackage{url}
\usepackage{verbatim}
\usepackage{graphicx}
\usepackage{cite}
\usepackage{multirow}
\usepackage{graphicx}
\usepackage{epstopdf}
\usepackage{stfloats}
\usepackage{subfigure}
\usepackage{microtype}
\usepackage{setspace}
\usepackage{color}
\usepackage{pifont}
\usepackage{xcolor}
\usepackage{bm}
\makeatletter
\renewcommand{\maketag@@@}[1]{\hbox{\m@th\normalsize\normalfont#1}}%
\makeatother

\begin{document}

\title{Geometry-Based Stochastic Probability Models for the LoS and NLoS Paths of A2G Channels \\ under Urban Scenarios}

\author{Minghui~Pang;
        Qiuming~Zhu,~\IEEEmembership{Member,~IEEE;}
        Cheng-Xiang Wang,~\IEEEmembership{Fellow,~IEEE;}
        Zhipeng Lin,~\IEEEmembership{Member,~IEEE;}
        Junyu Liu,~\IEEEmembership{Member,~IEEE;}
        Chongyu Lv,
        Zhuo Li,~\IEEEmembership{Member,~IEEE}
\thanks{This work was supported in part by the National Key Scientific Instrument and Equipment Development Project under Grant No. 61827801, in part by the open research fund of State Key Laboratory of Integrated Services Networks, No. ISN22-11, in part by Natural Science Foundation of Jiangsu Province, No. BK20211182, in part by the open research fund of National Mobile Communications Research Laboratory, Southeast University, No. 2022D04. \emph{(Corresponding author: Qiuming~Zhu and Cheng-Xiang Wang)}}
\thanks{M.~Pang, Q.~Zhu, Z.~Lin and C.~Lv are with the Key Laboratory of Dynamic Cognitive System of Electromagnetic Spectrum Space, College of Electronic and Information Engineering, Nanjing University of Aeronautics and Astronautics, Nanjing 211106, China. M.~Pang and Q.~Zhu are also with State Key Laboratory of Integrated Services Networks, Xidian University, Xian 710000, China (e-mail: \{pangminghui; zhuqiuming; linlzp\}@nuaa.edu.cn, lvcy2166@163.com).}
\thanks{C.-X.~Wang is with National Mobile Communications Research Laboratory, School of Information Science and Engineering, Southeast University, Nanjing 210096, China. He is also with Purple Mountain Laboratories, Nanjing 211111, China (e-mail: chxwang@seu.edu.cn).}
\thanks{J.~Liu is with State Key Laboratory of Integrated Services Networks, Xidian University, Xian 710000, China (e-mail: junyuliu@xidian.edu.cn).}
\thanks{Zhuo.~Li is with The Key Laboratory of Radar Imaging and Microwave Photonics, College of Electronic and Information Engineering, Nanjing University of Aeronautics and Astronautics, Nanjing 211106, China (e-mail: lizhuo@nuaa.edu.cn).}}

\markboth{IEEE INTERNET OF THINGS JOURNAL~Vol.~XX, No.~X, X~2022}%
{Shell \MakeLowercase{\textit{et al.}}: A Sample Article Using IEEEtran.cls for IEEE Journals}


\maketitle

\begin{abstract}
Path probability prediction is essential to describe the dynamic birth and death of propagation paths, and build the accurate channel model for air-to-ground (A2G) communications. The occurrence probability of each path is complex and time-variant due to fast changeable altitudes of UAVs and scattering environments. Considering the A2G channels under urban scenarios, this paper presents three novel stochastic probability models for the line-of-sight (LoS) path, ground specular (GS) path, and building-scattering (BS) path, respectively. By analyzing the geometric stochastic information of three-dimensional (3D) scattering environments, the proposed models are derived with respect to the width, height, and distribution of buildings. The effect of Fresnel zone and altitudes of transceivers are also taken into account. Simulation results show that the proposed LoS path probability model has good performance at different frequencies and altitudes, and is also consistent with existing models at the low or high altitude. Moreover, the proposed LoS and NLoS path probability models show good agreement with the ray-tracing (RT) simulation method.
\end{abstract}


\section{Introduction}
\IEEEPARstart{U}{nmanned} aerial vehicles (UAVs) have been widely used in aerial photography, disaster rescue, and other fields due to small size, low cost, and easy deployment \cite{Xiao20_ITJ,HeRS20_TVT,Cheng20_ITJ,Liu20_CC,Guizani20_trans,WCX21_SCIC}. The air-to-ground (A2G) communication technology provides important support for space-ground integrated communication networks \cite{Wang20_VTM,Li19_ITJ,Molisch20_TVT,Wang19_ICC,Zhu20_IWCMC,Rinaldi21_TB}. However, the dynamic birth and death of propagation paths needs to be carefully considered for A2G communication technology and channel modeling \cite{Khawaja19_CST,Cai19_TVT,Rappaport19_JSAC,Zhu18_trans,Liu21_IJSAC,Zhao16_CC,Mao20_Sensors,Huang21_TWC,ZhangJH20_CC}. Accurate and general propagation path probability models should be proposed, which can describe the dynamic process of birth and death and evaluate the system performance.
\par There are limited literature on the path probability prediction. Some researchers used accurate digital maps to determine the line-of-sight (LoS) path by geometric operation, namely deterministic method. This method is only suitable for a specific scenario and requires accurate maps. On the other hand, stochastic methods \cite{ITU-R2135,3GPP,5GCM,WINNER,Holis08_TAP,Samimi15_WCL,Lee18_TAP,Zhu21_FITEE,Jarvelainen16_WCL,ITU-R1410,Hourani14_WCL,Hourani20_WCL,Liu18_CL,Cui20_ITJ,Gapeyenko21_TWC} are more popular, which can be mainly classified into geometry-based analytical method \cite{ITU-R1410,Hourani14_WCL, Hourani20_WCL, Liu18_CL, Cui20_ITJ, Gapeyenko21_TWC}, measurement-based empirical method \cite{ITU-R2135,3GPP,5GCM,WINNER}, and simulation-based empirical method \cite{Holis08_TAP, Samimi15_WCL, Lee18_TAP, Zhu21_FITEE,Jarvelainen16_WCL}.
\par The measurement-based empirical method and simulation-based empirical method establish the stochastic path probability models by analyzing massive measured and simulated data. There are several standard LoS probability models based on the massive measurement data under several typical scenarios in International Telecommunication Union-Radio (ITU-R) M.2135-1 \cite{ITU-R2135}, the Third Generation Partnership Project (3GPP) TR 38.901 \cite{3GPP}, 5G Channel Model (5GCM) \cite{5GCM}, and WINNER II \cite{WINNER}. Since it is complex and high-cost to acquire measurement data, several empirical models are proposed based on simulation data, e.g., ray-tracing (RT) method \cite{Holis08_TAP, Samimi15_WCL, Lee18_TAP, Zhu21_FITEE}, and point cloud method \cite{Jarvelainen16_WCL}. Whereas, this kind of methods require huge computation for most applications.
\par The geometry-based analytical method predicts the path probability by applying the electromagnetic wave propagation theory and geometry information. Especially, the scenarios are described in a stochastic way according to the stochastic properties of buildings \cite{ITU-R1410,Hourani14_WCL, Hourani20_WCL, Liu18_CL, Cui20_ITJ, Gapeyenko21_TWC}. For example, a well-known analytical method was proposed in ITU-R Rec. P.1410 \cite{ITU-R1410}, which utilized three stochastic geometry parameters to define different urban scenarios. The buildings were distributed uniformly and the height of buildings followed Rayleigh distribution. On this basis, several LoS probability models were addressed in \cite{Hourani14_WCL, Hourani20_WCL, Liu18_CL, Cui20_ITJ, Gapeyenko21_TWC}. For example, the authors in \cite{Hourani14_WCL} proposed a LoS probability model with respect to the elevation angle under urban scenarios. However, the model was only appropriate for the high-altitude case (over 10 km). In \cite{Hourani20_WCL}, the authors described the buildings by cylinders, whose heights obeyed Log-normal distribution, and the building distribution followed Poisson point process. Furthermore, considering the factor of Fresnel zone, the authors in \cite{Liu18_CL,Cui20_ITJ} proposed frequency-dependent LoS probability models. In \cite{Gapeyenko21_TWC}, the authors proposed an altitude-dependent LoS probability model for UAV communications. Note that most of aforementioned models only consider the LoS probability, while the none-LoS (NLoS) path including the ground specular (GS) path and building scattering (BS) path are rarely involved. This paper aims to fill this gap and build the stochastic altitude-dependent and frequency-dependent probability models for different paths. The main novelties and contributions are summarized as follows:
\par 1) A general stochastic prediction model for the LoS probability of A2G channels in urban scenarios is proposed. This model takes the factors of transceiver altitude, building height, building width, building location, and the Fresnel zone into account, which makes it general and suitable for different altitudes and frequencies.
\par 2) Geometry-based stochastic probability models for the GS and BS paths in urban scenarios are proposed for the first time. We adopt the research methods of mirror scenario analysis and the first Fresnel zone restriction, and consider the effect of Fresnel ellipsoids and the geometric stochastic information such as transceiver locations and distribution of buildings. The two proposed models can be adapted to different heights, frequencies, and urban scenarios.
\par 3) Based on the stochastic scenario-dependent parameters, we construct virtual urban scenarios and obtain the LoS/GS/BS path probabilities by averaging massive RT simulation. The simulation results show that the proposed models have good agreement with the RT data. The proposed model is also consistent with existing models at low or high altitude. In addition, we analyze the influence factors and average maximum communication distance (MCD) based on the proposed path probability models.
\par The remainder of this paper is organized as follows. Section~II demonstrates the propagation process and analyzes the factors that affect the occurrence probability. In Section~III, the probability models of LoS path, GS path, and BS path are analyzed and derived for typical urban scenarios. The comparisons and validations are given in Section~IV. Finally, conclusions are drawn in Section~V.
\section{Propagation path of A2G channels}
The A2G channel generally consists of a LoS path and several NLoS paths, i.e., the GS paths and BS paths. In \cite{Mao20_Sensors}, the authors mentioned that the sum power of LoS path, GS path and BS path with the largest power exceeds 99\% of the total received power. Therefore, the channel impulse response (CIR) can be simplified as
\setcounter{equation}{0}
\begin{equation}
\sigma(\tau ,t) = {\sigma^{{\rm{LoS}}}}(t) + {\sigma^{{\rm{GS}}}}(\tau ,t) + {\sigma^{{\rm{BS}}}}(\tau ,t),
\label{1}
\end{equation}
\noindent where ${\sigma^{{\rm{LoS}}}}(t)$, ${\sigma^{{\rm{GS}}}}(\tau ,t)$, and ${\sigma^{{\rm{BS}}}}(\tau ,t)$ denote the CIRs of LoS, GS, and BS paths. Note that the occurrence of each path is random, which mainly depends on the scattering environment and locations of transceivers. This paper focuses on predicting the occurrence of each path in a stochastic way, denoted by the path probabilities as ${P^{{\rm{LoS}}}}$, ${P^{{\rm{GS}}}}$, and ${P^{{\rm{BS}}}}$, respectively.
\setcounter{figure}{0}
\begin{figure}[!t]
\centering
\subfigure[3D view of a typical A2G channel]{
\includegraphics[width=75mm]{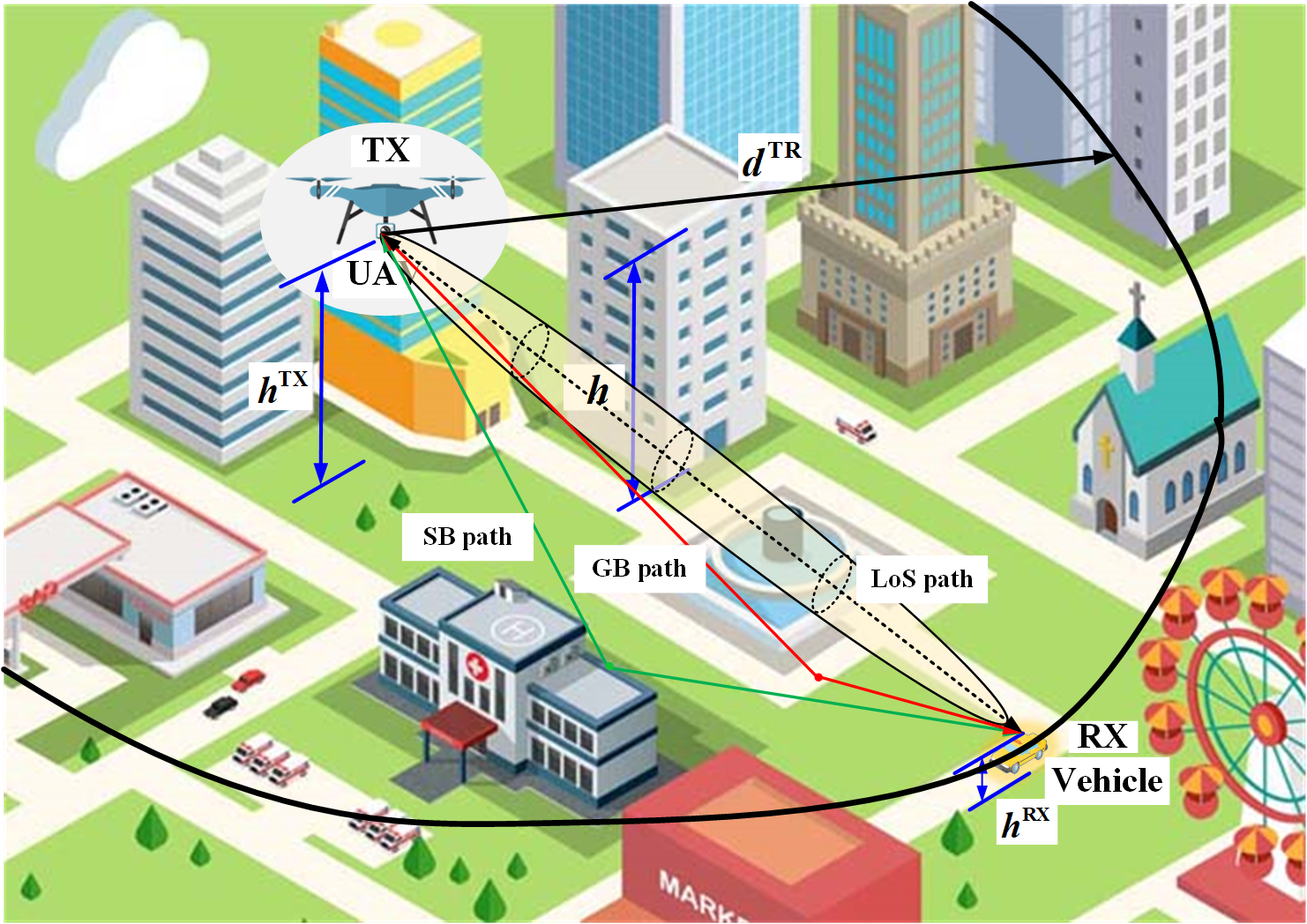}
}
\subfigure[2D view of a typical A2G channel]{
\includegraphics[width=75mm]{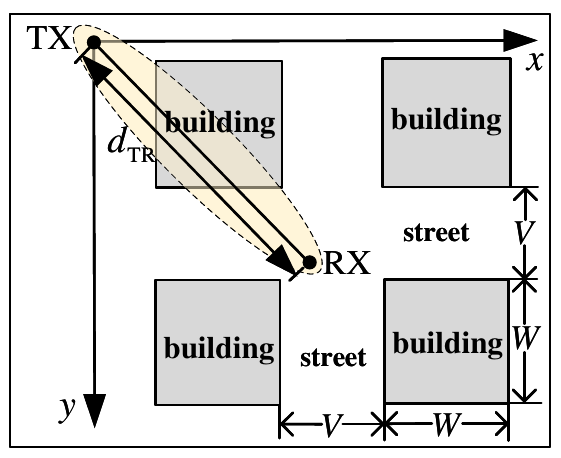}
}
\caption{An illustration of propagation paths of an A2G urban channel.}
 \label{fig:1}
\end{figure}
\par In order to build the probability model, we firstly need to describe the scattering scenario in a stochastic way. In this paper, we adopt the well-known classification and description method in \cite{Holis08_TAP} and \cite{ITU-R1410}. The method divides the urban scenarios into four typical categories, i.e., Suburban, Urban, Dense urban, and High-rise urban. Four typical categories are described by three parameters ${\rm{\{ }}\alpha ,\beta ,\gamma {\rm{\} }} \in \psi $, where $\alpha $ is the percentage of land area covered by buildings, $\beta $ represents the density of buildings, and $\gamma $ denotes the mean value of random building height. The location of buildings follows Uniform distribution and the height follows Rayleigh distribution as
\setcounter{equation}{1}
\begin{equation}
F(h) = \frac{h}{{{\gamma ^2}}}\exp \left[ { - \frac{{{{(h)}^2}}}{{2{\gamma ^2}}}} \right] ,
\label{2}
\end{equation}
\noindent where $h$ is the height of buildings.
\par A typical A2G channel under urban scenarios is shown in Fig.~1, where ${{h}^{\text{TX}}}$ and ${{h}^{\text{RX}}}$ represent the heights of transmitter (TX) and receiver (RX), respectively, and ${d^{{\rm{TR}}}}$ is the horizontal distance between TX and RX. In the figure, the average building width \cite{Hourani14_GCC} can be obtained by
\setcounter{equation}{2}
\begin{equation}
W = 1000\sqrt {\alpha /\beta },
\label{3}
\end{equation}
\setcounter{figure}{1}
\begin{figure*}[!t]
	\centering
	\includegraphics[width=120mm]{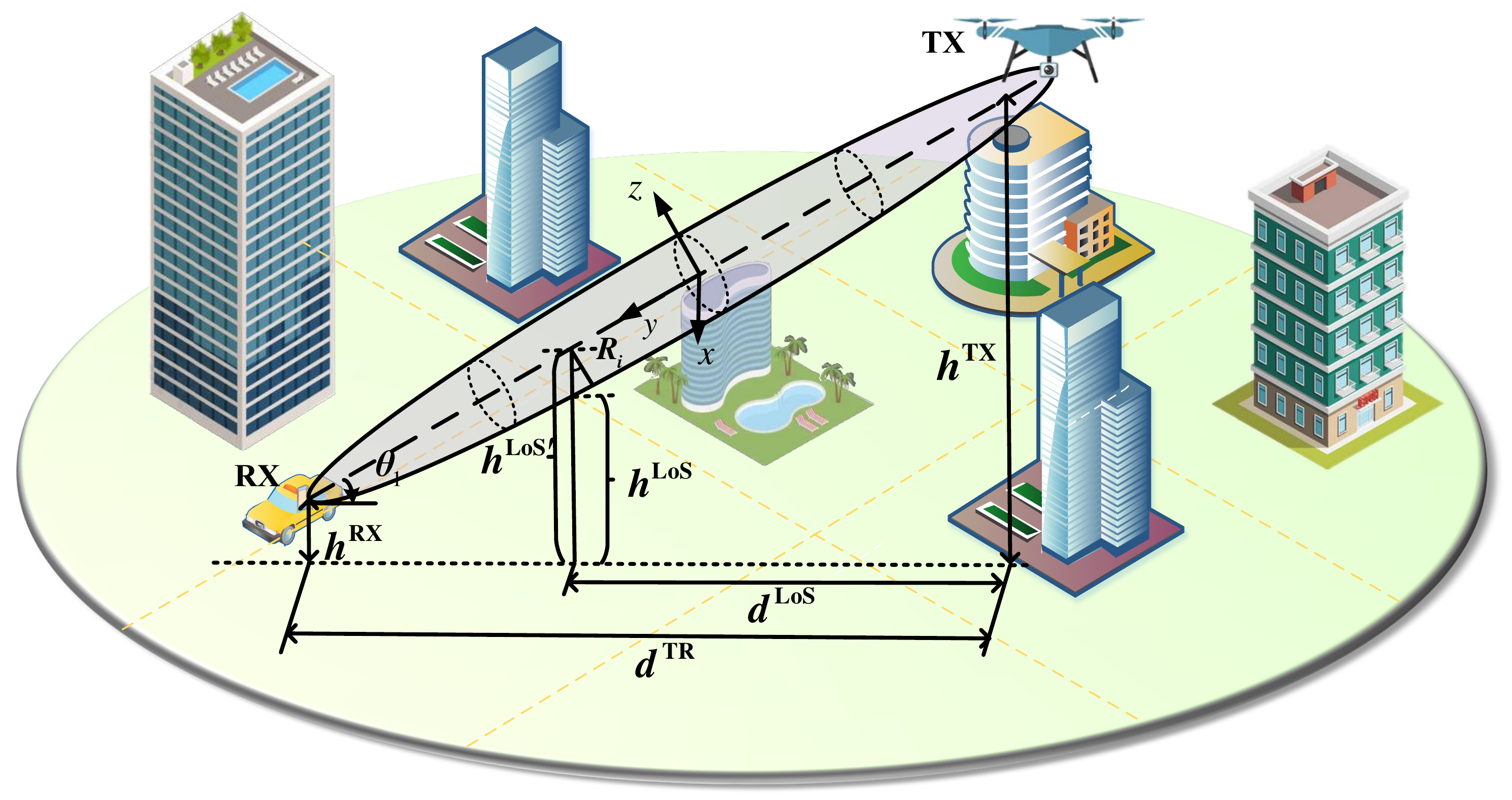}
	\caption{LoS path of A2G channel under urban scenarios.}
    \label{fig:2}
\end{figure*}
\hspace{-1.5mm}and the average street width can be obtained by
\setcounter{equation}{3}
\begin{equation}
V = 1000/\sqrt \beta   - W.
\label{4}
\end{equation}
The average number of buildings along the propagation path between the transceivers can be calculated by
\setcounter{equation}{4}
\begin{equation}
E[N] = {\rm{floor}}({d^{{\rm{TR}}}}\sqrt {\alpha \beta } /1000),
\label{5}
\end{equation}
where floor(.) represents the downward rounding function.
\par The energy of propagation path is not concentrated along a straight line, but in the entire Fresnel zone. When we consider the occurrence of propagation path, it is necessary to judge the occlusion of obstacles not only along the straight path but also on the entire Fresnel zone. The Fresnel clearance zone can be expressed as \cite{Liu18_CL, Cui20_ITJ}
\setcounter{equation}{5}
\begin{equation}
\frac{{{x^2}}}{{{X^2}}} + \frac{{{y^2}}}{{{Y^2}}} + \frac{{{z^2}}}{{{Z^2}}} \le 1,
\label{6}
\end{equation}
where the ellipsoid parameters can be calculated by
\setcounter{equation}{6}
\begin{equation}
\left\{ \begin{array}{l}
X = Z = \frac{{\sqrt {\lambda {d^{{\rm{TR}}}}} }}{2}\\
Y = \sqrt {\frac{{\lambda {d^{{\rm{TR}}}}}}{4} + \frac{{{{({d^{{\rm{TR}}}})}^2}}}{{\rm{4}}}}
\end{array} \right.,
\label{7}
\end{equation}
\begin{figure*}[!t]
\setcounter{equation}{14}
\begin{small}
\begin{equation}
{P_{{\rm{LoS}}}}({d^{{\rm{TR}}}},{h^{{\rm{TX}}}},{h^{{\rm{RX}}}},\psi ,\lambda ) = \prod\limits_{i = {\rm{1}}}^{E[N]} {\left[ {1 - \exp \left( { - \frac{{{{\left[ {{h^{{\rm{TX}}}} - \frac{{{d_i}({h^{{\rm{TX}}}} - {h^{{\rm{RX}}}})}}{{{d^{{\rm{TR}}}}}} - \frac{{\sqrt {\lambda {d^{{\rm{TR}}}}} \sqrt {{{({d^{{\rm{TR}}}})}^2} + {{({h^{{\rm{TX}}}} - {h^{{\rm{RX}}}})}^2}} }}{{{{({d^{{\rm{TR}}}})}^2}}}\min ({d_i},{d^{{\rm{TR}}}} - {d_i})} \right]}^2}}}{{2{\gamma ^{\rm{2}}}}}} \right)} \right]}
\label{15}
  \vspace{-3ex}
\end{equation}
\end{small}
\end{figure*}
\hspace{-2.0mm} where $\lambda $ is the wavelength, and $n$ is the order index of Fresnel zone. In this paper, we only consider the first-order Fresnel zone which includes half of total field strength. The radius of the first-order Fresnel ellipsoid corresponding to the location of the $i$-th building is
\setcounter{equation}{7}
\begin{equation}
R_i^{} = \left\{ \begin{array}{l}
\frac{{\sqrt {\lambda {d^{{\rm{TR}}}}} {d_i}}}{{{d^{{\rm{TR}}}}}},{d_i} \le \frac{{{d^{{\rm{TR}}}}}}{2}\\
\frac{{\sqrt {\lambda {d^{{\rm{TR}}}}} ({d^{{\rm{TR}}}} - {d_i})}}{{{d^{{\rm{TR}}}}}},{d_i} > \frac{{{d^{{\rm{TR}}}}}}{2}
\end{array} \right.,
\label{8}
\end{equation}
where ${d_i}$ is the distance between the $i$-th building and TX, given by
\setcounter{equation}{8}
\begin{equation}
{d_i} = \frac{{(i - 0.5){d^{{\rm{TR}}}}}}{{{\rm{floor(}}{d^{{\rm{TR}}}}\sqrt {\alpha \beta } /1000)}} + \frac{W}{2} .
\label{9}
\end{equation}
\section{Stochastic path probability Models}
\subsection{LoS Path Probability Model}
\setcounter{figure}{2}
\begin{figure*}[!b]
	\centering
	\includegraphics[width=130mm]{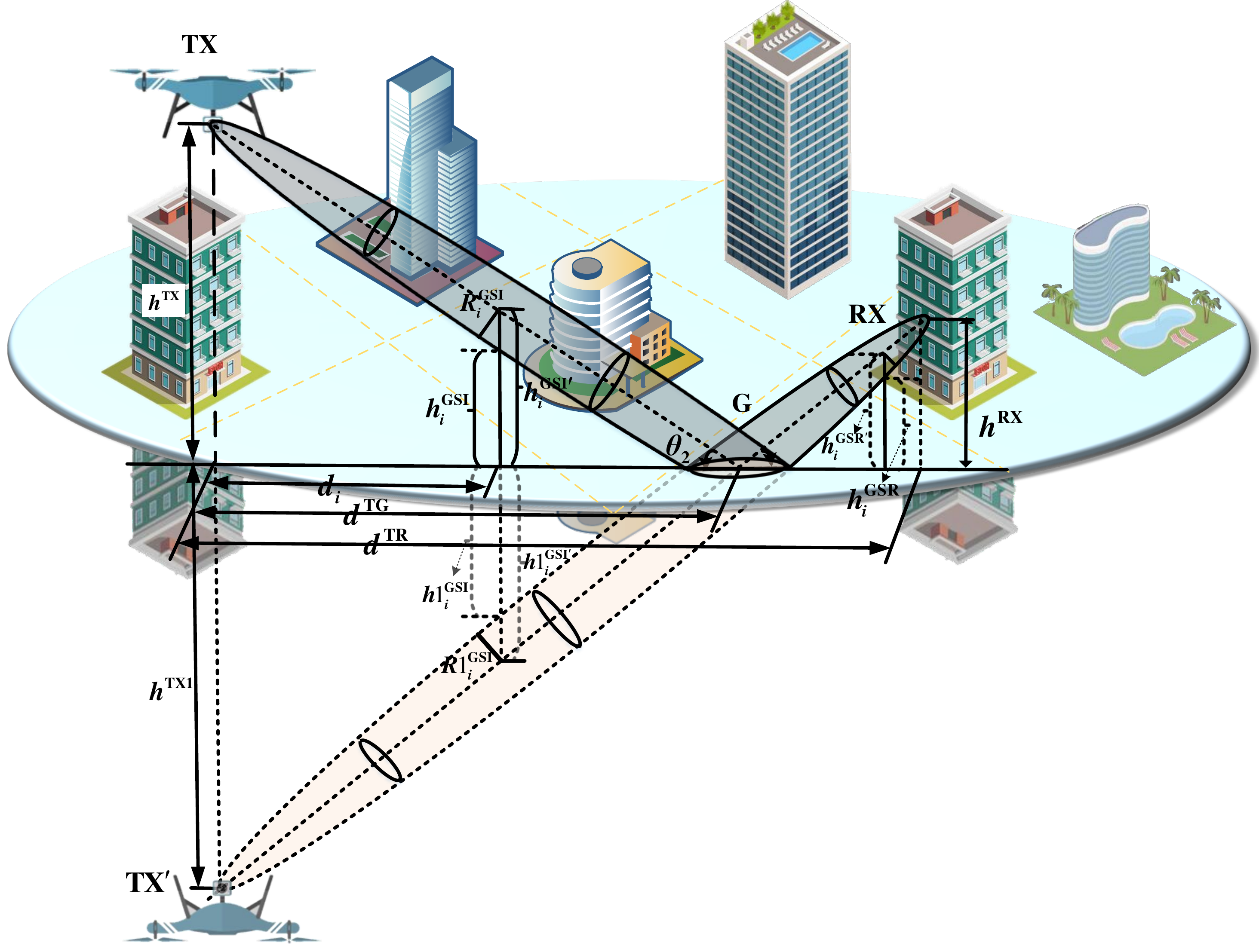}
	\caption{GS path of A2G channel under urban scenarios.}
    \label{fig:3}
\end{figure*}
\par Most of scatterers under A2G communication scenarios are on the ground and the LoS propagation is cut off only when the Fresnel zone is totally blocked by obstacle as shown in Fig.~2. In other word, the Fresnel zone between TX and RX is not blocked by any building. Thus, the LoS probability can be defined as
\setcounter{equation}{9}
\begin{equation}
{P^{{\rm{LoS}}}} = \prod\limits_{i = 1}^{E[N]} {P_i^{{\rm{LoS}}} = } \prod\limits_{i = 1}^{E[N]} {P({h_i} < h_i^{{\rm{LoS}}})},
\label{10}
\end{equation}
where $P_i^{{\rm{LoS}}}$ is the probability that the $i$-th building does not block the LoS path, ${h_i}$ is the height of the $i$-th building, and $h_i^{{\rm{LoS}}}$ is the maximum height of the $i$-th building unblocking the LoS path.
\par Since the height of buildings follows the Rayleigh distribution, the probability that the buildings do not block the LoS path is
\setcounter{equation}{10}
\begin{small}
\begin{equation}
P_i^{{\rm{LoS}}} = P({h_i} < h_i^{{\rm{LoS}}}) = \int_0^{h_i^{{\rm{LoS}}}} {F(h){\rm{d}}h = 1 - \exp } \left[ { - \frac{{{{(h_i^{{\rm{LoS}}})}^2}}}{{2{\gamma ^2}}}} \right].
\label{11}
\end{equation}
\end{small}
\hspace{-3.3mm} When the effect of Fresnel ellipsoid is not considered, the maximum height point is along the straight line between TX and RX, which can be expressed as
\setcounter{equation}{11}
\begin{equation}
h{_i^{{{\rm{LoS}}}^\prime}} = {h^{{\rm{TX}}}} - \frac{{{d_i}({h^{{\rm{TX}}}} - {h^{{\rm{RX}}}})}}{{{d^{{\rm{TR}}}}}} .
\label{12}
\end{equation}
The influence of Fresnel ellipsoid depending on the signal frequency is considered in this paper. The maximum height can be modified as the lowest point of first-order Fresnel zone as shown in Fig.~2. It yields
\setcounter{equation}{12}
\begin{equation}
h{_i^{{\rm{LoS}}}} = {h^{{\rm{TX}}}} - \frac{{{d_i}({h^{{\rm{TX}}}} - {h^{{\rm{RX}}}})}}{{{d^{{\rm{TR}}}}}} - \frac{{R_i}}{{\cos {\theta _1}}} .
\label{13}
\end{equation}
Furthermore, we can obtain
\setcounter{equation}{13}
\begin{equation}
\cos {\theta _1} = \frac{{{d^{{\rm{TR}}}}}}{{\sqrt {{{({d^{{\rm{TR}}}})}^2} + {{({h^{{\rm{TX}}}} - {h^{{\rm{RX}}}})}^2}} }}.
\label{14}
\end{equation}
Finally, the LoS probability can be obtained by Eq. (15), which is related with the altitudes of TX and RX, the communication frequency, and the scenario.
\subsection{GS Path Probability Model}
\par The GS propagation path consists of three parts, i.e., the incident path, reflection path, and Fresnel reflection zone. The location and height of buildings obey the aforementioned distributions. As shown in Fig.~3, the area under the ground is the mirror of the counterpart above the ground. The area that the Fresnel ellipsoid between mirrored ${\rm{TX'}}$ and RX crosses the ground can be denoted by G, i.e., the reflection Fresnel zone. Then, the probability that the $i$-th building does not block the GS path can be equivalent to the LoS probability between RX and mirrored ${\rm{TX'}}$ as shown in Fig.~3. Thus, it can be expressed as
\setcounter{equation}{15}
\begin{equation}
P_i^{{\rm{GS}}} = P({h_i} < h_i^{{\rm{GS}}}) = \int_0^{h_i^{{\rm{GS}}}} {F(h){\rm{d}}h = 1 - \exp } \left[ { - \frac{{{{(h_i^{{\rm{GS}}})}^2}}}{{2{\gamma ^2}}}} \right].
\label{16}
\end{equation}
\par However, it is slightly different from calculating LoS probability, since the buildings are divided into two classes according to their positions in this paper. Therefore, the GS path exists only when the building between TX and G does not block the incident path, and the building between G and RX does not block the reflection path. Note that the case of reflection zone G not blocked by any building is already included in above two constraints. The distance between G and TX can be obtained by performing the triangle similarity theorem as
\setcounter{equation}{16}
\begin{equation}
{d^{{\rm{TG}}}} = \frac{{{d^{{\rm{TR}}}}{h^{{\rm{TX}}}}}}{{{h^{{\rm{TX}}}} + {h^{{\rm{RX}}}}}} .
\label{17}
\end{equation}
Then, the cosine of ${\theta _2}$ in Fig.~3 can be obtained as
\setcounter{equation}{17}
\begin{equation}
\cos {\theta _2} = \frac{{{d^{{\rm{TR}}}}}}{{\sqrt {{{({d^{{\rm{TR}}}})}^2} + {{({h^{{\rm{TX}}}} + {h^{{\rm{RX}}}})}^2}} }} .
\label{18}
\end{equation}
\setcounter{equation}{22}
\begin{figure*}[!b]
\begin{equation}
P_i^{{\rm{GSI}}}({d^{{\rm{TR}}}},{h^{{\rm{TX}}}},{h^{{\rm{RX}}}},\psi ,\lambda ) = 1 - \exp \left( { - \frac{{{{\left[ {\frac{{{h^{{\rm{TX}}}}({d^{{\rm{TG}}}} - {d_i})}}{{{d^{{\rm{TG}}}}}} - \frac{{\sqrt {n\lambda {d^{{\rm{TR}}}}} \sqrt {{{({d^{{\rm{TR}}}})}^2} + {{({h^{{\rm{TX}}}} + {h^{{\rm{RX}}}})}^2}} }}{{{{({d^{{\rm{TR}}}})}^2}}}\min ({d_i},{d^{{\rm{TR}}}} - {d_i})} \right]}^2}}}{{2{\gamma ^{\rm{2}}}}}} \right)
\label{23}
\end{equation}
\end{figure*}
\begin{figure*}[!b]
\setcounter{equation}{27}
\begin{equation}
P_i^{{\rm{GSR}}}({d^{{\rm{TR}}}},{h^{{\rm{TX}}}},{h^{{\rm{RX}}}},\psi ,\lambda ) = 1 - \exp \left( { - \frac{{{{\left[ {\frac{{{h^{{\rm{RX}}}}\left( {{d_i} - {d^{{\rm{TG}}}}} \right)}}{{{d^{{\rm{TR}}}} - {d^{{\rm{TG}}}}}} - \frac{{\sqrt {n\lambda {d^{{\rm{TR}}}}} \sqrt {{{({d^{{\rm{TR}}}})}^2} + {{({h^{{\rm{TX}}}} + {h^{{\rm{RX}}}})}^2}} }}{{{{({d^{{\rm{TR}}}})}^2}}}\min ({d_i},{d^{{\rm{TR}}}} - {d_i})} \right]}^2}}}{{2{\gamma ^{\rm{2}}}}}} \right)
	\label{28}
\end{equation}
\end{figure*}
\setcounter{figure}{3}
\begin{figure*}[!b]
	\centering
	\includegraphics[width=120mm]{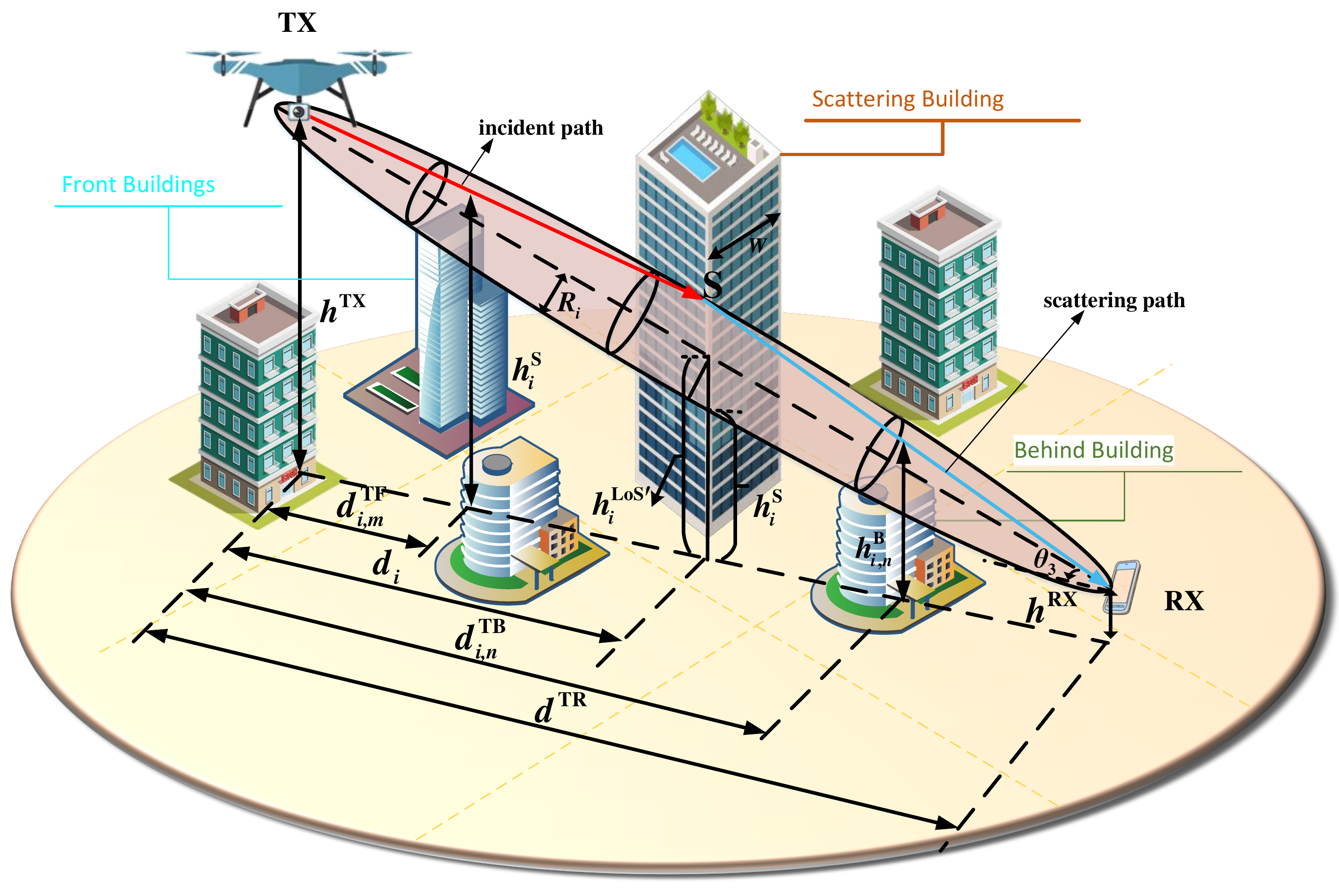}
	\caption{BS path of A2G channel under urban scenarios.}
    \label{fig:4}
\end{figure*}
\par Firstly, we assume that the arbitrary building is located between the TX and G, and then, the average number of buildings along the incident path can be calculated by
\setcounter{equation}{18}
\begin{equation}
E[{N^{\rm{I}}}] = {\rm{floor}}({d^{{\rm{TG}}}}\sqrt {\alpha \beta } /1000) .
\label{19}
\end{equation}
The distance ${{d}_{i}}$ between TX and the $i$-th building can be expressed as
\setcounter{equation}{19}
\begin{equation}
{d_i} < {d^{{\rm{TG}}}} = \frac{{{d^{{\rm{TR}}}}{h^{{\rm{TX}}}}}}{{{h^{{\rm{RX}}}} + {h^{{\rm{TX}}}}}} .
\label{20}
\end{equation}
The height of the centerline of incident Fresnel ellipsoid $h{{_{i}^{\text{GS}}}^{\prime }}$ is the same as the height when the incident path is considered as a straight line. It can be given by
\setcounter{equation}{20}
\begin{equation}
h{_i^{{{\rm{GSI}}}^\prime}} = \frac{{{h^{{\rm{TX}}}}\left( {{d^{{\rm{TG}}}} - {d_i}} \right)}}{{{d^{{\rm{TG}}}}}} .
\label{21}
\end{equation}
Considering the influence of Fresnel zone, the maximum height at which the $i$-th building does not block the incident path is
\setcounter{equation}{21}
\begin{equation}
h_i^{{\rm{GSI}}} = h{_i^{{{\rm{GSI}}}^\prime }} - \frac{{R_i^{}}}{{\cos {\theta _2}}} .
\label{22}
\end{equation}
The probability of the $i$-th building does not block the incident ray can be derived as Eq. (23).
\par Secondly, if the building is located between the RX and G, the average number of buildings along the reflection path can be calculated by
\setcounter{equation}{23}
	\begin{equation}
    E[{{N}^{\text{R}}}]=\text{floor}(({{d}^{\text{TR}}}-{{d}^{\text{TG}}})\sqrt{\alpha \beta }/1000) .
	\label{24}
	\end{equation}
The distance ${d_i}$  between the TX and the $i$-th building can be expressed as
\setcounter{equation}{24}
\begin{equation}
{{d}_{i}}>{{d}^{\text{TG}}}=\frac{{{d}^{\text{TR}}}{{h}^{\text{TX}}}}{{{h}^{\text{RX}}}+{{h}^{\text{TX}}}} .
	\label{25}
\end{equation}
Similarly, we can obtain
\setcounter{equation}{25}
\begin{equation}
	h{{_{i}^{\text{GSR}}}^{\prime }}=\frac{{{h}^{\text{RX}}}\left( {{d}_{i}}-{{d}^{\text{TG}}} \right)}{{{d}^{\text{TR}}}-{{d}^{\text{TG}}}},
	\label{26}
\end{equation}
and
\setcounter{equation}{26}
\begin{equation}
h_i^{{\rm{GSR}}} = h{_i^{{{\rm{GSR}}}^\prime} } - \frac{{R_i^{}}}{{\cos {\theta _2}}} .
	\label{27}
\end{equation}
Then, the probability that the $i$-th building does not block the reflection path can be obtained as Eq. (28).
Thus, the final probability model of GS path can be obtained as
\setcounter{equation}{28}
\begin{equation}
	{{P}^{\text{GS}}}({{d}^{\text{TR}}},{{h}^{\text{TX}}},{{h}^{\text{RX}}},\psi ,\lambda )=\prod\limits_{i=\text{1}}^{E[{{N}^{\text{I}}}]}{P_{i}^{\text{GSI}}}\prod\limits_{i=\text{1}}^{E[{{N}^{\text{R}}}]}{P_{i}^{\text{GSR}}} .
	\label{29}
\end{equation}
It should be noted that the case of buildings on the reflection Fresnel zone is included in Eq. (29). Since buildings are usually much larger than the reflection Fresnel zone, they would block the incident path or reflection path when close to the reflection zone.
\subsection{BS Path Probability Model}
\setcounter{equation}{35}
\begin{figure*}[!b]
\begin{equation}
P_i^{\rm{S}}({d^{{\rm{TR}}}},{h^{{\rm{TX}}}},{h^{{\rm{RX}}}},\psi ,\lambda ) = \exp \left( { - \frac{{{{\left[ {{h^{{\rm{TX}}}} - \frac{{d_i^{}({h^{{\rm{TX}}}} - {h^{{\rm{RX}}}})}}{{{d^{{\rm{TR}}}}}} - \frac{{\sqrt {\lambda {d^{{\rm{TR}}}}} \sqrt {{{({d^{{\rm{TR}}}})}^2} + {{({h^{{\rm{TX}}}} - {h^{{\rm{RX}}}})}^2}} }}{{{{({d^{{\rm{TR}}}})}^2}}}\min (d_i^{},{d^{{\rm{TR}}}} - d_i^{})} \right]}^2}}}{{2{\gamma ^{\rm{2}}}}}} \right)
	\label{36}
 \vspace{-3ex}
\end{equation}
\end{figure*}
\par Since the double-bounce and multiple-bounce scattering paths have much lower power compared with the single-bounce scattering path, they are not considered in this paper. Moreover, any one building can become a scattering zone when the height of the building enters the first Fresnel zone. A typical BS path in A2G communications is shown in Fig.~4. According to the position of scattering zone, the buildings along the straight line can be classified into three categories, e.g., scattering buildings, front buildings, and behind buildings. Note that the scattering occurs on the scattering buildings. The front buildings are located between TX and the scattering zone, and the behind buildings are located between RX and the scattering zone.
\par During the scattering propagation process, it is necessary to ensure that the incident and scattering buildings do not block the incident path and the scattering path. Therefore, the BS probability model can be expressed as
\setcounter{equation}{29}
\begin{small}
\begin{equation}
	{{P}^{\text{BS}}}=1-\prod\limits_{i=1}^{E[N]}{\left[ 1-P_{i}^{\text{S}}\cdot \left( \prod\limits_{m=1}^{E[{{N}^{\text{F}}}]}{P_{i,m}^{\text{F}}} \right)\cdot \left( \prod\limits_{n=1}^{E[{{N}^{\text{B}}}]}{P_{i,n}^{\text{B}}} \right) \right]} ,
	\label{30}
\end{equation}
\end{small}
\hspace{-1.3mm}where $E[{N^{\rm{F}}}]$ is the average number of front buildings. When the scattering zone is on the $i$-th building, it can be expressed as
\setcounter{equation}{30}
\begin{equation}
	E[{{N}^{\text{F}}}]=\text{floor(}d_{i}^{{}}\sqrt{\alpha \beta }/1000) .
	\label{31}
\end{equation}
The average number of behind buildings can be obtained as
\setcounter{equation}{31}
\begin{equation}
	E[{{N}^{\text{B}}}]=\text{floor}\left[ ({{d}^{\text{TR}}}-d_{i}^{{}})\sqrt{\alpha \beta }/1000 \right] .
	\label{32}
\end{equation}
Then, the probability that the $i$-th building enters the Fresnel zone can be derived as
\setcounter{equation}{32}
\begin{small}
\begin{equation}
	P_{i}^{\text{S}}=P({{h}_{i}}>h_{i}^{\text{S}})=1-\int_{0}^{h_{i}^{\text{S}}}{F(h)\text{d}h=\exp }\left[ -\frac{{{(h_{i}^{\text{S}})}^{2}}}{2{{\gamma }^{2}}} \right] .
	\label{33}
\end{equation}
\end{small}
\hspace{-1.3mm}According to Eq. (13) and Eq. (14), the minimum height of the $n$-th building entering the first Fresnel ellipsoid is
\setcounter{equation}{33}
\begin{equation}
	h_{i}^{\text{S}}={{h}^{\text{TX}}}-\frac{d_{i}^{{}}({{h}^{\text{TX}}}-{{h}^{\text{RX}}})}{{{h}^{\text{RX}}}}-\frac{{{R}_{i}}}{\cos {{\theta }_{3}}} .
	\label{34}
\end{equation}
According to triangular geometry, the cosine of ${\theta _3}$  in Fig.~4 can be expressed as
\setcounter{equation}{34}
\begin{equation}
	\cos {{\theta }_{3}}=\frac{{{d}^{\text{TR}}}}{\sqrt{{{({{d}^{\text{TR}}})}^{2}}+{{({{h}^{\text{TX}}}-{{h}^{\text{RX}}})}^{2}}}} .
	\label{35}
\end{equation}
Therefore, the probability that the $i$-th building height enters the Fresnel ellipsoid can be calculated as Eq. (36).
\par The probability that the $m$-th front building does not block the incident path can be expressed as
\setcounter{equation}{36}
\begin{small}
\begin{equation}
	P_{i,m}^{\text{F}}=P({{h}_{i,m}}<h_{i,m}^{\text{F}})=\int_{0}^{h_{i,m}^{\text{F}}}{F(h)\text{d}h=1-\exp }\left[ -\frac{{{(h_{i,m}^{\text{F}})}^{2}}}{2{{\gamma }^{2}}} \right] .
	\label{37}
\end{equation}
\end{small}
\hspace{-1.3mm}Here, the maximum height of the $m$-th front building $h_{i,m}^{\rm{F}}$ can be expressed as
\setcounter{equation}{37}
\begin{equation}
	h_{i,m}^{\text{F}}=\frac{(d_{i}^{{}}-d_{i,m}^{\text{TF}})({{h}^{\text{TX}}}-h_{i}^{\text{S}})}{d_{i}^{{}}}+h_{i}^{\text{S}} ,
	\label{38}
\end{equation}
where $d_{i,m}^{{\rm{TF}}}$ is the horizon distance between TX and the $m$-th front building,
\setcounter{equation}{38}
\begin{equation}
	d_{i,m}^{\text{TF}}=\frac{(i-0.5)d_{i}^{{}}}{\text{floor}\left( d_{i}^{{}}\sqrt{\alpha \beta }/1000 \right)}+\frac{W}{2} .
	\label{39}
\end{equation}
Then, the probability that the $m$-th front building does not block incident ray can be obtained as
\setcounter{equation}{39}
\begin{scriptsize}
\begin{equation}
	P_{i,m}^{\rm{F}}({d^{{\rm{TR}}}},{h^{{\rm{TX}}}},{h^{{\rm{RX}}}},\psi ,\lambda ) = 1 - \exp \left( { - \frac{{{{\left[ {\frac{{(d_i^{} - d_{i,m}^{{\rm{TF}}})({h^{{\rm{TX}}}} - h_i^{\rm{S}})}}{{d_i^{}}} + h_i^{\rm{S}}} \right]}^2}}}{{2{\gamma ^2}}}} \right) .
	\label{40}
\end{equation}
\end{scriptsize}
\par The probability that the $n$-th behind building does not block the scattering path can be defined as
\setcounter{equation}{40}
\begin{small}
\begin{equation}
	P_{i,n}^{\text{B}}=P({{h}_{i,n}}<h_{i,n}^{\text{B}})=\int_{0}^{h_{i,n}^{\text{B}}}{F(h)\text{d}h=1-\exp }\left[ -\frac{{{(h_{i,n}^{\text{B}})}^{2}}}{2{{\gamma }^{2}}} \right] ,
	\label{41}
\end{equation}
\end{small}
\hspace{-1.5mm}where $h_{i,n}^{\rm{B}}$ denotes the maximum height of the $n$-th front building. Similarly, we can obtain
\setcounter{equation}{41}
\begin{scriptsize}
\begin{equation}
\hspace{-1.5mm}
	P_{i,n}^{\rm{B}}({d^{{\rm{TR}}}},{h^{{\rm{TX}}}},{h^{{\rm{RX}}}},\psi ,\lambda ) = 1 - \exp \left( { - \frac{{{{\left[ {{h^{{\rm{RX}}}} + \frac{{({d^{{\rm{TR}}}} - d_{i,n}^{{\rm{TB}}})({h^{{\rm{TX}}}} - h_i^{\rm{S}})}}{{{d^{{\rm{TR}}}} - d_i^{}}}} \right]}^2}}}{{2{\gamma ^2}}}} \right) ,
	\label{42}
\end{equation}
\end{scriptsize}
\hspace{-1.5mm}where the horizon distance between TX and the $n$-th front building can be expressed as
\setcounter{equation}{42}
\begin{equation}
	d_{i,n}^{\text{TB}}=\frac{(i-0.5)d_{i}^{{}}}{\text{floor}\left( d_{i}^{{}}\sqrt{\alpha \beta }/1000 \right)}+\frac{W}{2} .
	\label{43}
\end{equation}
By substituting (31)--(33), (40) and (43) into (30), we can obtain the BS probability model of single bounce.
\setcounter{figure}{4}
\begin{figure}[!b]
	\centering
	\includegraphics[width=80mm]{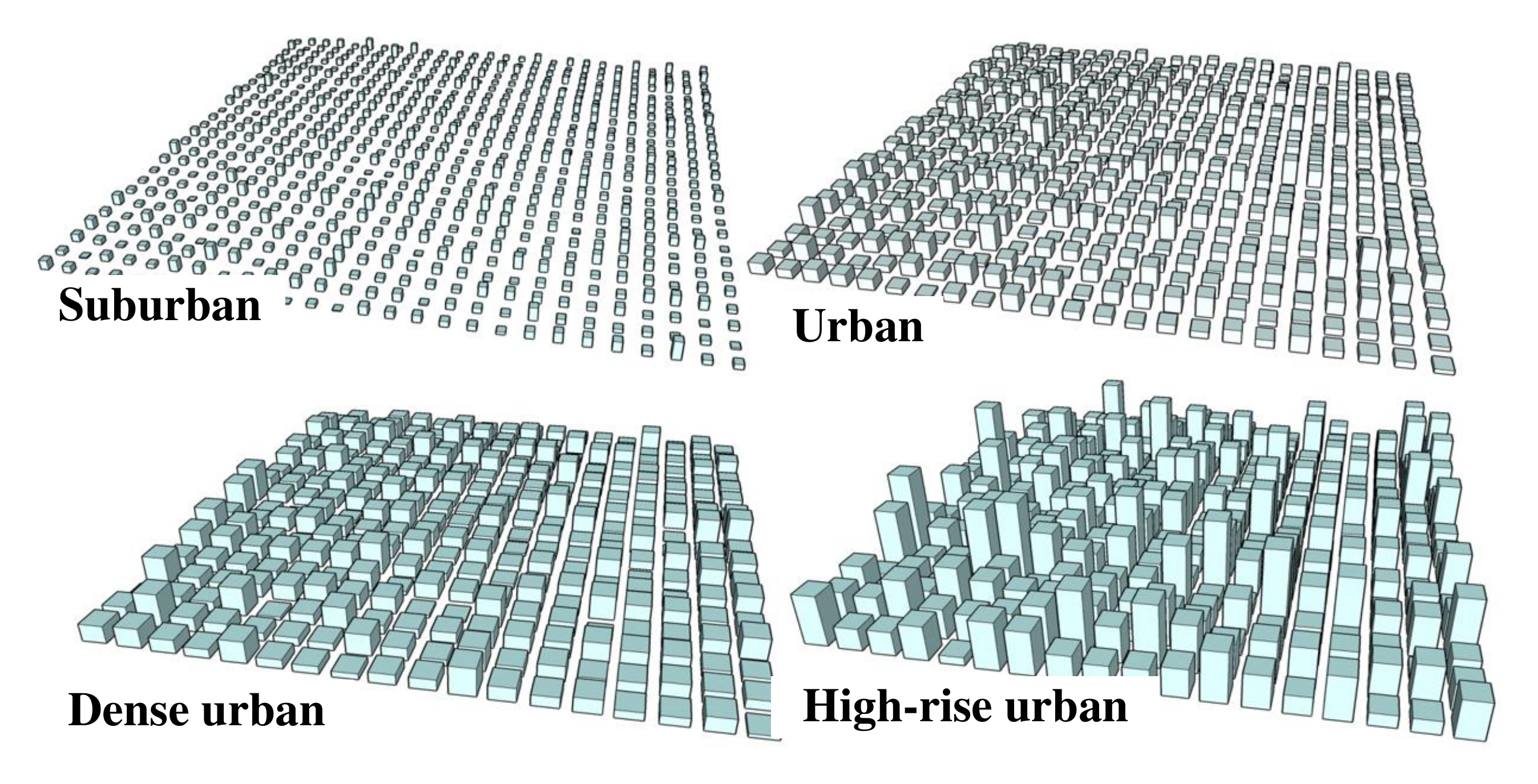}
	\caption{Four constructed urban scenarios.}
    \label{fig:5}
\end{figure}
\begin{table*}[!t]
\renewcommand{\arraystretch}{1.3}
\caption{RT Simulation parameters.}
\label{table_1}
\centering
\begin{tabular}{p{4.2cm}p{7.4cm}}
\hline
Parameter & Value\\
\hline
Scenario & Suburban, Urban, Dense urban and High-rise urban\\
Building area ratio ($\alpha $) & 0.1, 0.3, 0.5, 0.5\\
Number of buildings ($\beta $) & 750/${\rm{km}}^2$, 500/${\rm{km}}^2$, 300/${\rm{km}}^2$, 300/${\rm{km}}^2$\\
Building location distribution & Uniform distribution\\
Building width & 11.6 m, 24.5 m, 40.8 m, 40.8 m\\
Street width ($V$) & 24.9 m, 20.2 m, 16.9 m, 16.9 m\\
Building height distribution & Rayleigh distribution\\
Average building height ($\gamma $) & 8 m, 15 m, 20 m, 50 m\\
Frequency ($f$) & 28 GHz, 1.4 GHz\\
Antenna type & Omnidirectional\\
TX height (interval) & 0--1000 m (20 m)\\
TX number & 50\\
RX height & 2 m\\
RX number & 6825\\
\hline
\end{tabular}
\end{table*}
\section{Simulation results and validation}
\subsection{RT-based simulation method}
\par This section demonstrates the effectiveness and accuracy of proposed stochastic probability models. We conduct massive RT simulations under urban scenarios for quantitative comparison. In order to obtain the average RT results, we first construct the propagation scenarios defined by ITU-R \cite{ITU-R1410}, including standard Suburban, Urban, Dense urban, and High-rise urban, as shown in Fig.~5. The buildings and streets follow Uniform distribution. The random height of buildings follows Rayleigh distribution. Moreover, 50 TXs with the height from 10~m to 1000~m are evenly placed, and 6825 RXs at the altitude of 2~m are uniformly distributed on several concentric circles in each scenario, which covers most of possible local surrounding conditions. The detailed parameters are given in Table~I.
\par We apply the RT technique to the constructed scenarios and obtain the conditions of LoS, GS, and BS paths. The occurrence number of LoS, GS, and BS paths at the same altitude and distance can be obtained respectively. Then, the path probabilities of LoS, GS, and BS paths refer to the ratio of occurrence number to the total number.
\subsection{Comparison and Validation}
\par We first analyze and verify the LoS probability model which is described as a function of elevation angle at the high latitude (over 10 km) , as presented in \cite{Holis08_TAP} and \cite{Hourani14_WCL}. In Fig.~2, the elevation angle is defined as $\theta {\rm{  =  }}\arctan ({h^{{\rm{TR}}}}/{d^{{\rm{TR}}}})$ and the proposed LoS probability model can be transformed to the function of $\theta $. Set $f$ = 28~GHz, ${h^{{\rm{TX}}}}$ = 5005~m, ${h^{{\rm{RX}}}}$ = 5~m. The comparison results with respect to $\theta $ for different scenarios are shown in Fig.~6(a). It can be seen that the prediction trend of the proposed model is the same as that of the models in \cite{Holis08_TAP,Hourani14_WCL}.
\setcounter{figure}{5}
\begin{figure}[!t]
\centering
\subfigure[LoS probability VS elevation angle ($f$~=~28 GHz, ${h^{{\rm{TR}}}}$~=~5000~m)]{
\includegraphics[width=75mm]{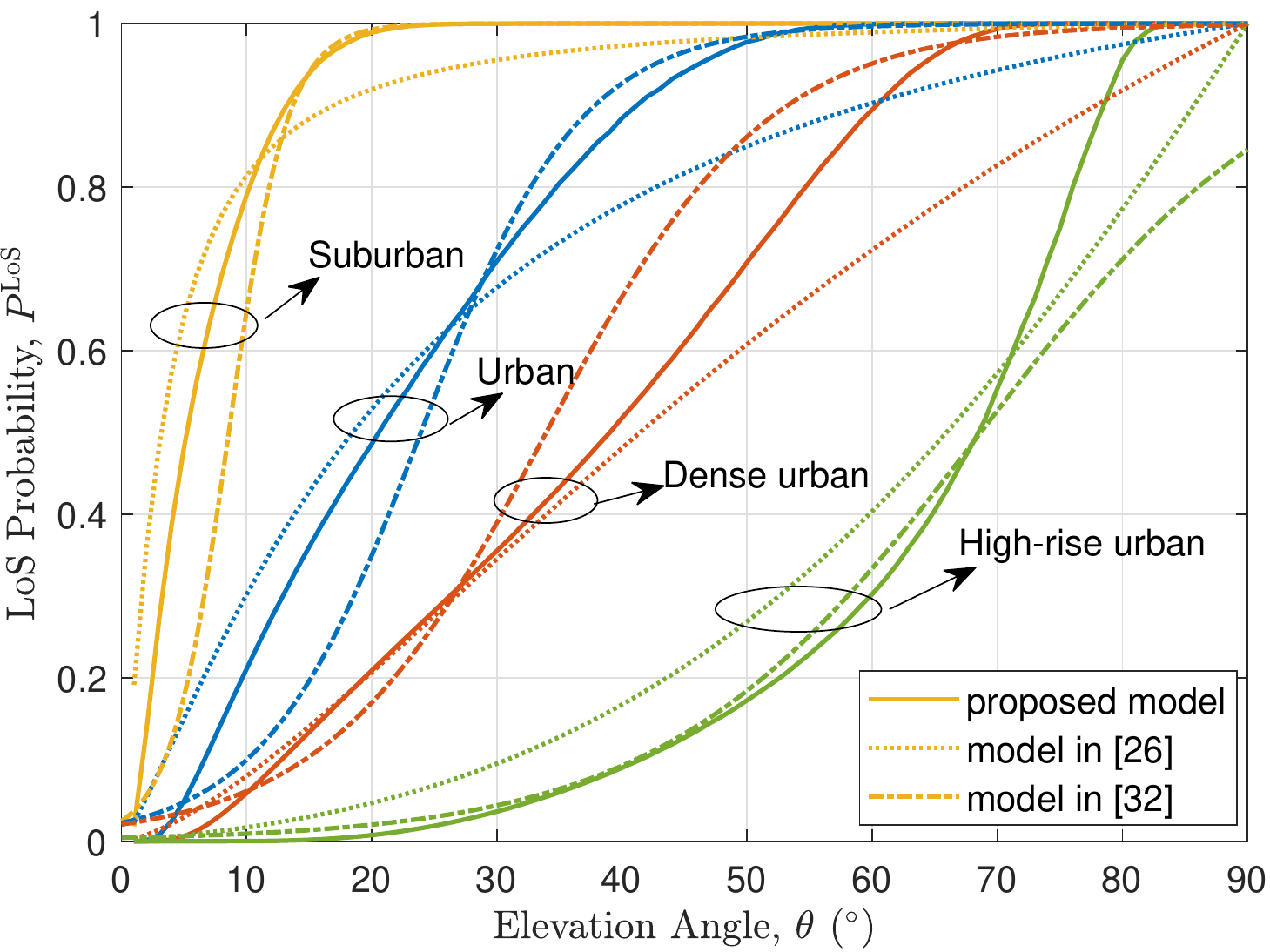}
}
\subfigure[LoS probability VS distance ($f$~=~28~GHz, ${h^{{\rm{TR}}}}$~=~30, 120, 500~m, Urban scenario)]{
\includegraphics[width=75mm]{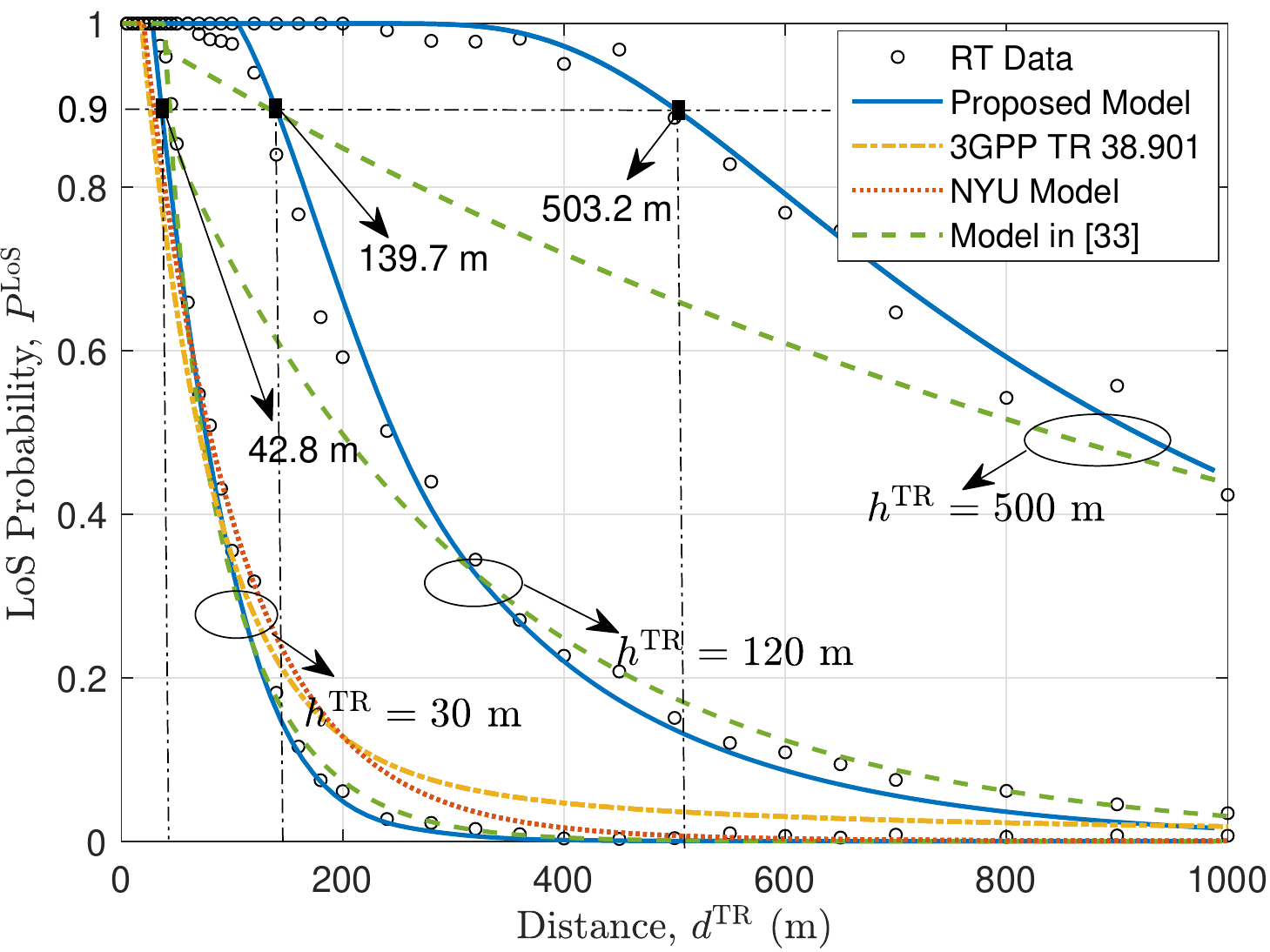}
}
\caption{Comparison of proposed LoS probability model and RT data.}
 \label{fig:6}
   \vspace{-2ex}
\end{figure}
\setcounter{figure}{6}
\begin{figure*}[!t]
\centering
\subfigure[LoS probability VS elevation angle]{
\includegraphics[width=75mm]{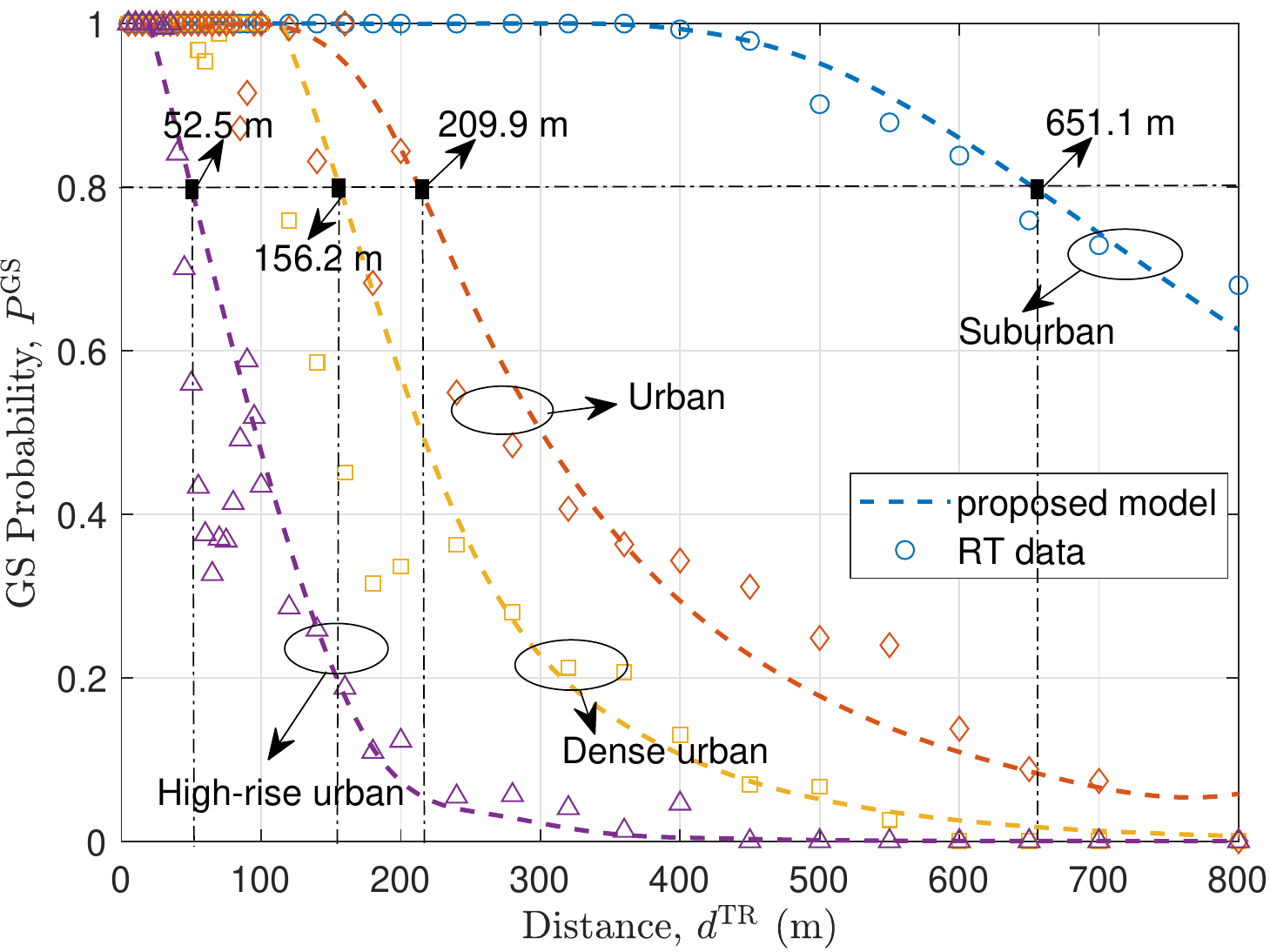}
}
\subfigure[LoS probability VS distance ]{
\includegraphics[width=75mm]{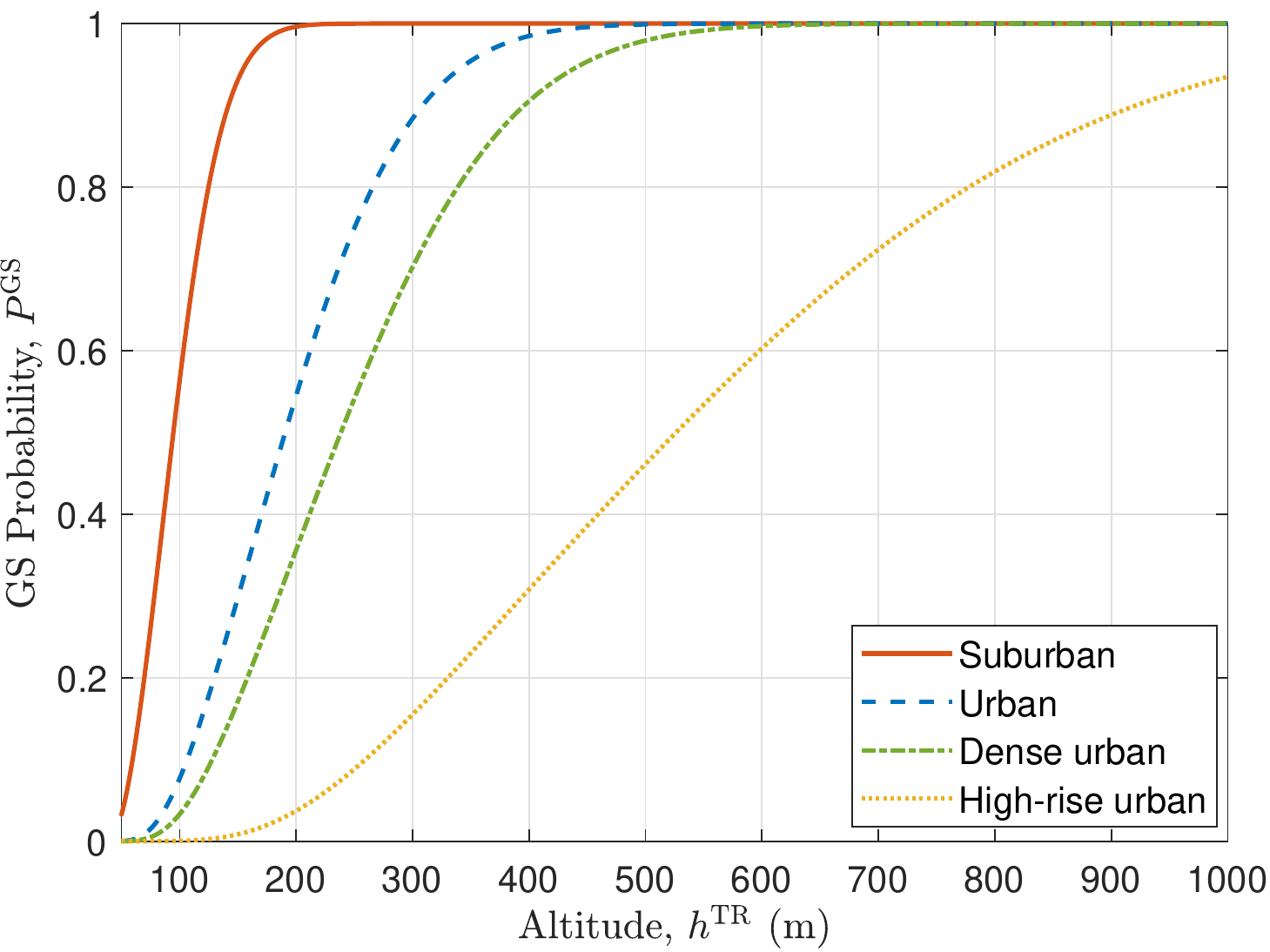}
}
\caption{Comparison of proposed GS probability model and RT data ($f$~=~1.4~GHz, ${h^{{\rm{TR}}}}$ = 200 m).}
 \label{fig:7}
    \vspace{-3ex}
\end{figure*}
\setcounter{figure}{7}
\begin{figure*}[!b]
\centering
\subfigure[Suburban]{
\includegraphics[width=75mm]{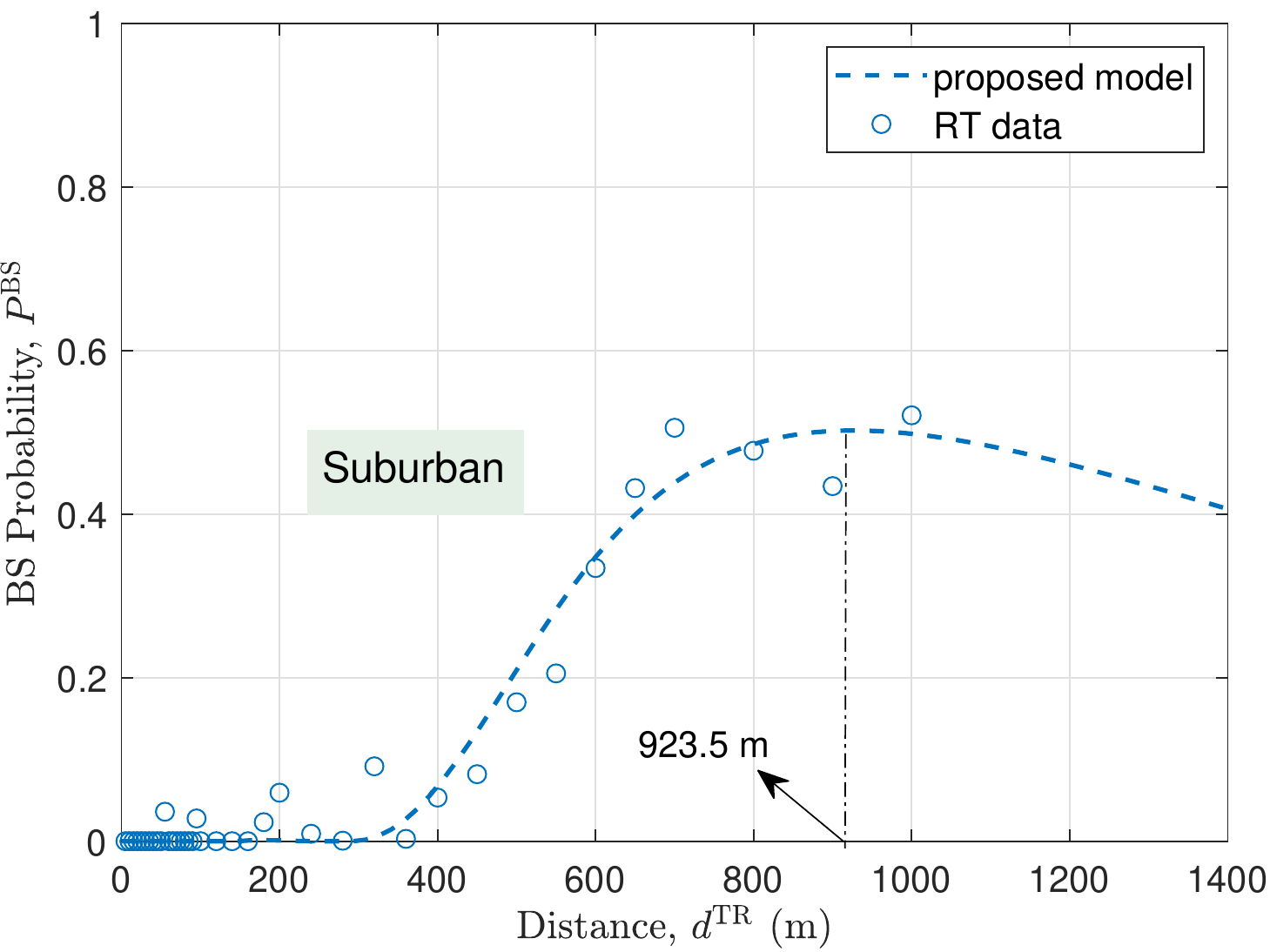}
}
\subfigure[Urban]{
\includegraphics[width=75mm]{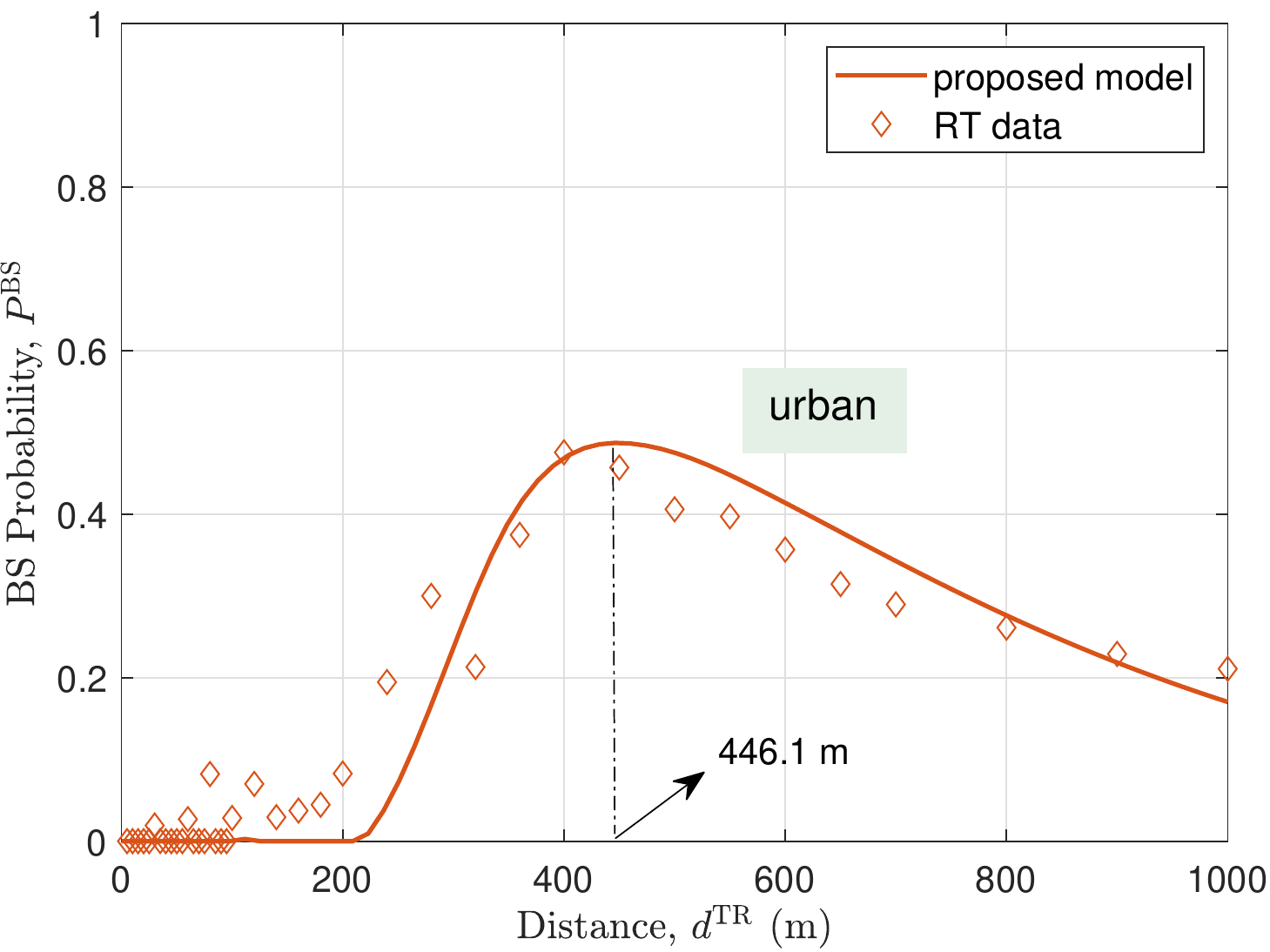}
}\\
\subfigure[Dense urban]{
\includegraphics[width=75mm]{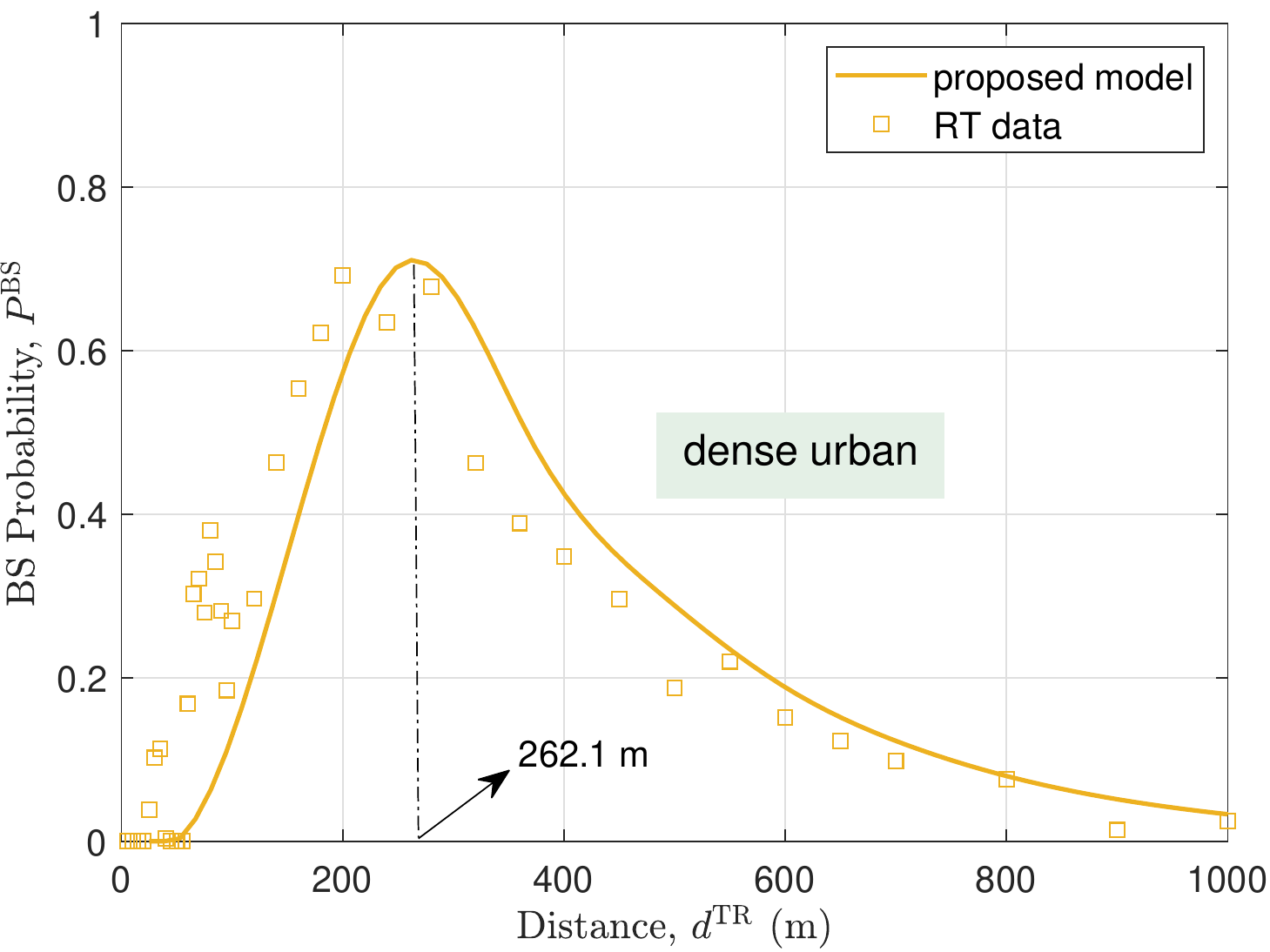}
}
\subfigure[High-rise urban]{
\includegraphics[width=75mm]{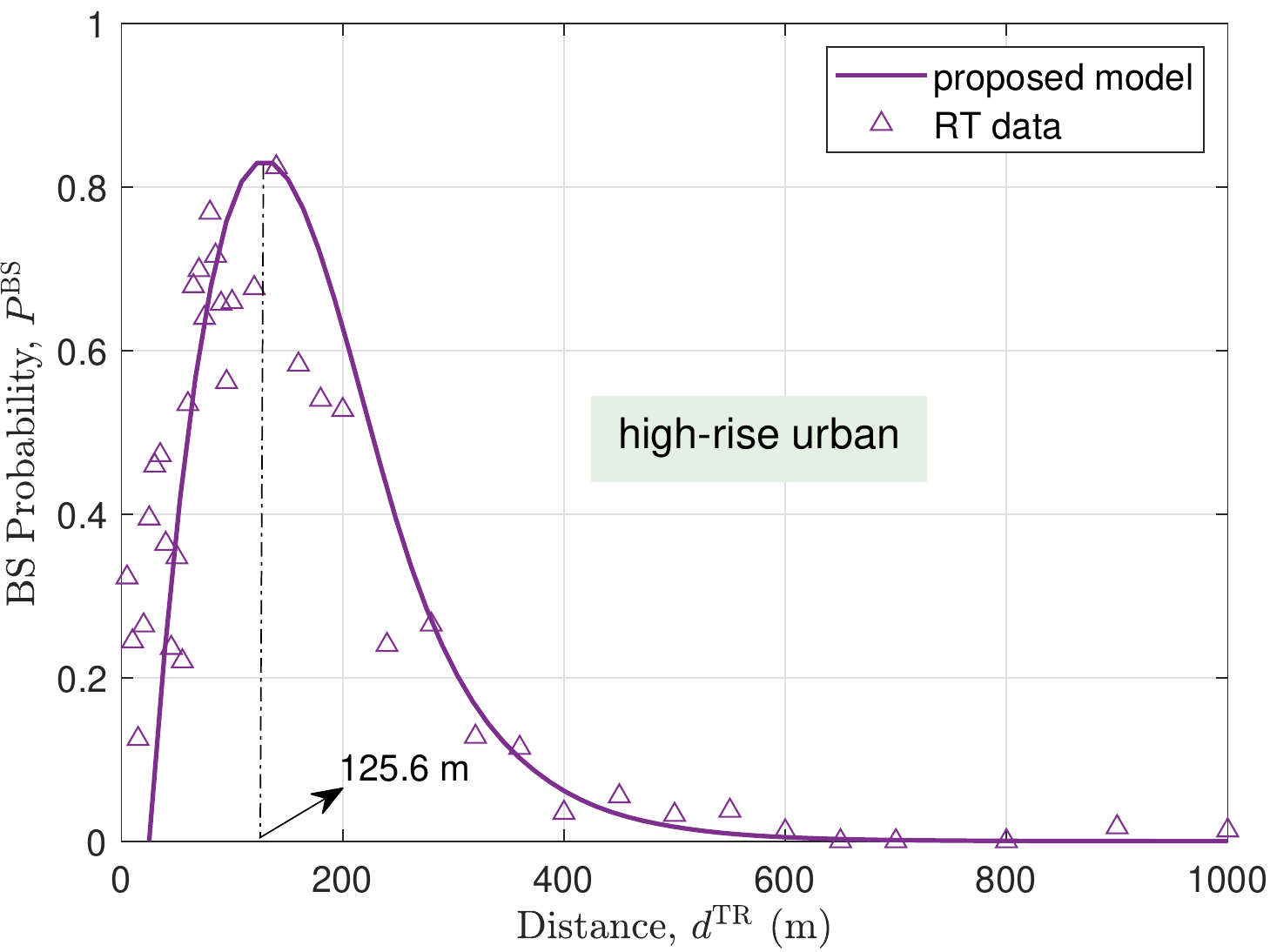}
}
\caption{Comparison of proposed BS probability model and RT data ($f$ = 1.4 GHz, ${h^{{\rm{TR}}}}$ = 200 m).}
 \label{fig:8}
\end{figure*}
\par The comparison of proposed LoS probability model, RT simulation method and the standard models under Urban scenario is shown in Fig.~6(b). Since the standard models are designed for low altitude scenarios, the relative height ${h^{{\rm{TR}}}} = {h^{{\rm{TX}}}} - {h^{{\rm{RX}}}}$ is set as 30~m. The model in \cite{Hourani14_WCL} is an altitude-dependent model but assumes the buildings of Poisson Point Process (PPP) distribution, so we set the relative height 120 m and 500 m. As we can see, the proposed LoS probability model shows good agreement with the standard models and the RT data at low altitude. As the altitude increases, the standard models are no longer applicable. However, our proposed model is still applicable and shows good agreement with the RT data. Note that LoS path is essential for the reliability of millimeter wave (mmWave) A2G communications. Taking the LoS probability of 0.9 as an example, we can obtain the average MCDs as 42.8~m, 139.7~m, and 503.2~m for different altitudes of UAV. This is because the LoS path are blocked by less buildings as the altitude increases. These quantitative results can be used to evaluate the cell coverage and placement of aerial base stations.
\par To the best of our knowledge, there are no probability models for the GS and BS paths of A2G channels. To verify the proposed GS probability model, we use the RT results under aforementioned four urban scenarios. Since the GS and BS paths have more influence on the sub-6G A2G communications, the simulation parameters are set as $f$ = 1.4~GHz, ${h^{{\rm{TX}}}}$ = 202~m, and ${h^{{\rm{RX}}}}$ = 2~m. As shown in Fig.~7(a), excellent agreement between the proposed model and RT data verifies the prediction accuracy. Moreover, in the four urban scenarios, the GS path probability gradually decreases with increasing communication distance since longer communication distance involves larger number of buildings which in turn increases the probability of occlusion. Taking the GS probability of 0.8 as an example, we can obtain the average MCDs as 651.1~m, 209.9~m, 156.2~m, and 52.5~m for different scenarios. It is shown that the High-rise urban is the most severely occluded scenario for A2G communication due to the occlusion of high buildings. Furthermore, we plot the relationship between the GS probability and communication altitude under different scenarios in Fig.~7(b). The GS probability rises as the altitude increases. The reason is that the increase in communication height makes less occlusion of the building height.
\setcounter{figure}{8}
\begin{figure*}[!b]
\centering
\subfigure[Suburban]{
\includegraphics[width=75mm]{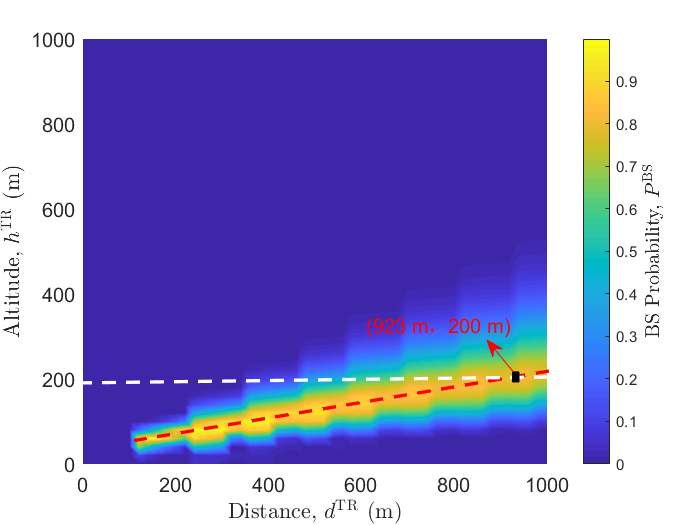}
}
\subfigure[Urban]{
\includegraphics[width=75mm]{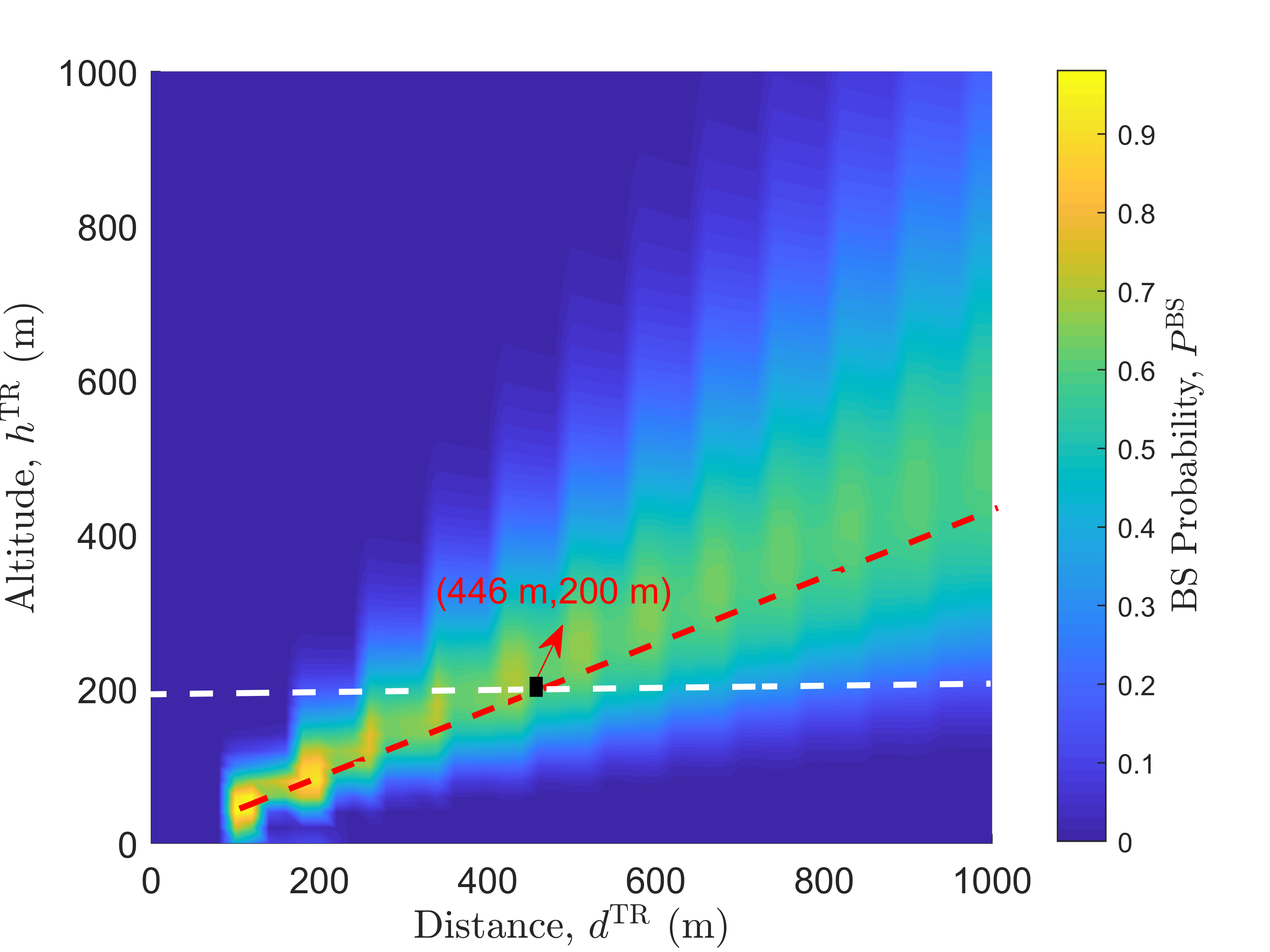}
}
\subfigure[Dense urban]{
\includegraphics[width=75mm]{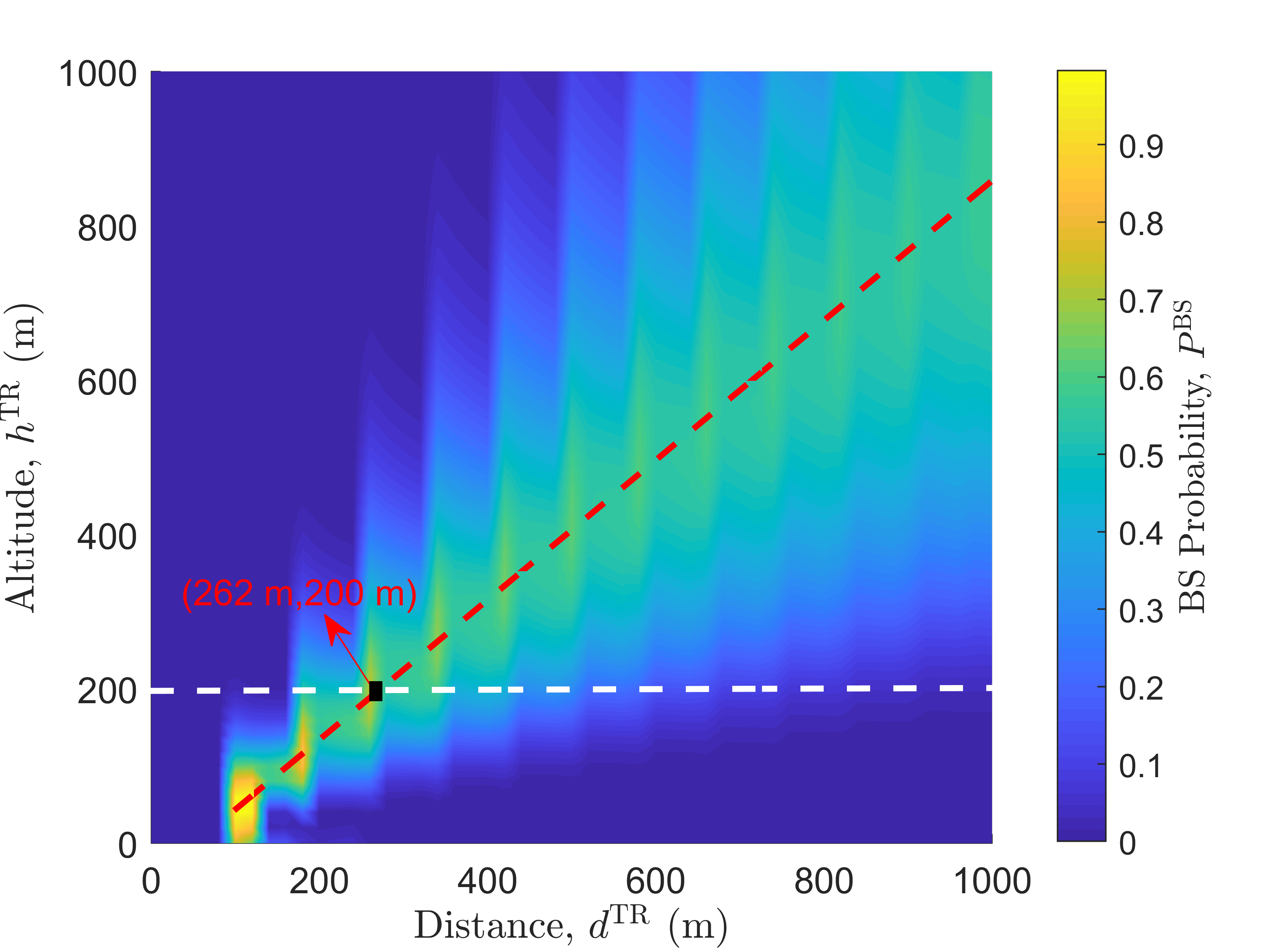}
}
\subfigure[High-rise urban]{
\includegraphics[width=75mm]{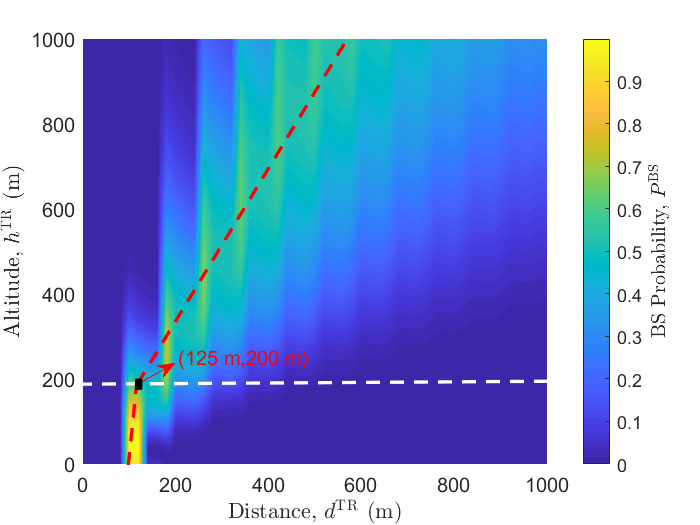}
}
\caption{BS path probabilities with different distances and altitudes ($f$ = 1.4 GHz, ${h^{{\rm{TR}}}}$ = 200 m).}
 \label{fig:9}
\end{figure*}
\par To verify the proposed BS probability model, we set $f$ = 1.4~GHz, ${h^{{\rm{TX}}}}$ = 202~m, ${h^{{\rm{RX}}}}$ = 2~m and compare the proposed model with the RT data under four urban scenarios in Fig.~8. We can see that the BS probability increases firstly and then decreases with the increasing distance. It is shown in Eq.~(30) that the trend of BS probability is mainly influenced by three factors $P_i^{\rm{S}}$, $P_{i,m}^{\rm{F}}$, and $P_{i,n}^{\rm{B}}$. For the rising stage, $P_i^{\rm{S}}$, depending on the scattering building, plays a major role on the BS probability. As the distance increases, the radius of Fresnel ellipsoid increases. Therefore, the scattering building is easy to enter the first Fresnel zone, and it makes the value of $P_i^{\rm{S}}$ rise. For the decreasing stage, the point whether the scattering propagation is obstructed by the front and behind buildings refers to $P_{i,m}^{\rm{F}}$ and $P_{i,n}^{\rm{B}}$. The long distance makes the probability $P_{i,m}^{\rm{F}}$ and $P_{i,n}^{\rm{B}}$ increase significantly. Moreover, as the scenario transforms from Suburban to High-rise urban, the optimal communication distances are 975.8~m, 470.9~m, 262.5~m, and 129.6~m, respectively. This is because the BS path is easy to be blocked when the density of buildings increases.
\par The prediction results are given in Fig.~9, which demonstrates the effects of altitude on the BS probability. As we can see, scenarios with dense and high buildings, such as the Dense urban and High-rise urban, the BS probability is large when the altitude is high. However, it is difficult to carry out long-distance communication. On the contrary, Suburban and Urban support long-distance communication but are not suitable for the communication with extremely high communication altitudes. Moreover, taking the communication altitude of 200~m as an example, we can easily get the MCDs under four scenarios from Fig.~9 as 930~m, 446~m, 262~m, 125~m, respectively, which can also be found in Fig.~8.
\section{Conclusions}
\par In this paper, we have proposed novel stochastic probability models that can predict the occurrence of LoS, GS, and BS paths in A2G communications. These models are based on the stochastic geometric information, i.e., the building height, the building width, and the building locations. The influence of transceiver altitudes, communication distance, and Fresnel ellipsoid zone have also been considered. These models have good versatility of different altitudes and frequencies. Simulation results have shown that the prediction results of LoS, GS, and BS path probabilities are consistent with the RT simulation data and also compatible with the existing probability models for the specific purpose. The research content can be used in enhancing the communication performance, signal coverage, and layout optimization of aerial base stations in the A2G integrated networks.

\end{document}